\documentclass[prl,aps,twocolumn,superscriptaddress]{revtex4-1}
\usepackage{bm}
\usepackage{graphicx}
\usepackage{amsmath}
\usepackage{amssymb}
\usepackage[utf8]{inputenc}
\usepackage[T1]{fontenc}
\usepackage{color}
\usepackage{upgreek}
\usepackage{subfigure}
\usepackage[unicode=true,colorlinks=true,citecolor=blue]{hyperref}
\usepackage{placeins}
\usepackage{lipsum}
\usepackage{epsfig}
\usepackage{makecell}
\usepackage[utf8]{inputenc}
\usepackage{wrapfig}
\usepackage{siunitx}
\newcommand{\nix}[1]{}
\usepackage{mhchem}
\usepackage{siunitx}

\newcommand{\CdXHgOneMinusXTe}{\ce{Cd_{x}Hg_{1-x}Te}~}

\newcommand{\CdHgTe}{\ce{CdHgTe}~}
\newcommand{\ZnTe}{\ce{ZnTn}~}
\newcommand{\CdTe}{\ce{CdTe}~}

\newcommand{\MilliMeter}[1]{\SI{#1}{\milli\metre}}
\newcommand{\MilliMeterSquared}[1]{\SI{#1}{\milli\metre\squared}}
\newcommand{\MicroMeter}[1]{\SI{#1}{\micro\meter}}
\newcommand{\NanoMeter}[1]{\SI{#1}{\nano\meter}}

\newcommand{\MilliWatt}[1]{\SI{#1}{\milli\watt}}

\newcommand{\KiloWatt}[1]{\SI{#1}{\kilo\watt}}
\newcommand{\Hertz}[1]{\SI{#1}{\hertz}}
\newcommand{\TeraHertz}[1]{\SI{#1}{\tera\hertz}}
\newcommand{\NanoSecond}[1]{\SI{#1}{\nano\second}}
\newcommand{\MicroAmperePerWatt}[1]{\SI[per-mode=symbol]{#1}{\micro\ampere\per\watt}}
\newcommand{\NanoAmperePerWatt}[1]{\SI[per-mode=symbol]{#1}{\nano\ampere\per\watt}}
\newcommand{\Degree}[1]{\SI{#1}{\degree}}

\begin{document}

    \definecolor{orange}{rgb}{1,0.5,0}

    \title{Symmetry breaking and circular photogalvanic effect in epitaxial \CdXHgOneMinusXTe films}
    \author{S. Hubmann}
    \affiliation{Terahertz Center, University of Regensburg, 93040 Regensburg, Germany}
     \author{G.\,V. Budkin}
    \affiliation{Ioffe Institute, 194021	St. Petersburg, Russia}
    \author{M. Otteneder}
    \affiliation{Terahertz Center, University of Regensburg, 93040 Regensburg, Germany}
    \author{D. But}
    \affiliation{International Research Centre CENTERA, Institute of High Pressure Physics, Polish Academy of Sciences PL-02346 Warsaw, Poland}
    \author{D. Sacré}
    \affiliation{Terahertz Center, University of Regensburg, 93040 Regensburg, Germany}
    \author{I. Yahniuk}
    \affiliation{International Research Centre CENTERA, Institute of High Pressure Physics, Polish Academy of Sciences PL-02346 Warsaw, Poland}
    \author{K. Diendorfer}
    \affiliation{Terahertz Center, University of Regensburg, 93040 Regensburg, Germany}
    \author{V.\,V. Bel'kov}
    \affiliation{Ioffe Institute, 194021	St. Petersburg, Russia}
    \author{D.\,A. Kozlov}
    \author{N.\,N. Mikhailov}
    \author{S.\,A. Dvoretsky}
    \author{V.\,S. Varavin}
    \author{V.\,G. Remesnik}
    \affiliation{Rzhanov Institute of Semiconductor Physics, 630090 Novosibirsk, Russia}
    \author{S.\,A. Tarasenko}
    \affiliation{Ioffe Institute, 194021	St. Petersburg, Russia}
    \author{W. Knap}
    \affiliation{International Research Centre CENTERA, Institute of High Pressure Physics, Polish Academy of Sciences PL-02346 Warsaw, Poland}
    \author{S.\,D. Ganichev}
    \affiliation{Terahertz Center, University of Regensburg, 93040 Regensburg, Germany}

    \begin{abstract}

          We report on the observation of symmetry breaking and the circular photogalvanic effect in \CdXHgOneMinusXTe alloys. We demonstrate that irradiation of bulk epitaxial films with circularly polarized terahertz radiation leads to the circular photogalvanic effect (CPGE) yielding a photocurrent whose direction reverses upon switching the photon helicity. This effect is forbidden in bulk zinc-blende crystals by symmetry arguments, therefore, its observation indicates either the symmetry reduction  of bulk material or that the photocurrent is excited in the topological surface states formed in a material with low Cadmium concentration. We show that the bulk states play a crucial role because the CPGE was also clearly detected in samples with non-inverted band structure. We suggest  that strain is a reason of the symmetry reduction. We develop a theory of the CPGE showing that the photocurrent results from  the quantum interference of different pathways contributing to the free-carrier absorption (Drude-like) of monochromatic radiation.

    \end{abstract}

    \maketitle

    \section{Introduction}
        \label{introduction}

        \CdXHgOneMinusXTe alloy, also known as MCT (Mercury Cadmium Telluride), is one of the most leading materials used for sensitive and fast infrared detectors \cite{Capper1997,norton2002,Henini2002,Rogalski2005,Downs2013,Rogalski2018,Vanamala2019}. Wide bandgap tunability of these materials allows radiation detection in an extremely wide frequency range, spanning from near- to mid-infrared wavelength. Furthermore, it has been used for the development of terahertz (THz) radiation detection, see e.g. Refs.~\cite{Dvoretsky2010,Rumyantsev2017,Ruffenach2017,Yavorskiy2018,Bak2018}. The introduction of the concept of topological insulators (TIs)~\cite{Moore2010,Hasan2010,Zhang2011} attracted great attention to novel aspects of \CdXHgOneMinusXTe compounds as well as low dimensional quantum heterostructures  based on these materials.
        The reason for that is the inverted band structure in HgTe and \CdXHgOneMinusXTe with a Cadmium concentration not exceeding the critical value $x_c$~\cite{Berchenko1976,Orlita2014,Malcolm2015,Teppe2016,Tomaka2017}, which is a crucial condition for the formation of helical surface or edge states~\cite{Zhang2011,Dyakonov1981,Tomaka2017,Marchewka2017,Minkov1996,Kane2005, Bernevig2006, Hasan2010, Zhang2011}.
        In comparison to other materials with a non-trivial band structure, CdHgTe-based compounds seem to be more promising due to their very high carrier mobility and feasibility to suppress effects from three-dimensional carriers in the sample volume.
        This is also supported by the well developed technological process originally motivated by the fabrication of detectors which has been adapted for the growth of TI materials on demand.
        The observation of Kane fermions in bulk \CdXHgOneMinusXTe crystals~\cite{Orlita2014,Malcolm2015,Teppe2016}, quantum spin Hall effect~\cite{Bernevig2006,Koenig2007,Nowack2013} and helical edge photocurrents in CdHgTe/HgTe/CdHgTe quantum wells~\cite{Dantscher2017} as well as the demonstration of Dirac surface states in 3D TI made of strained HgTe~\cite{Bruene2011,Shuvaev2012,Kozlov2014,Bruene2014,Dantscher2015}, are some
        important achievements in the physics of topological insulators. In the  case of bulk \CdXHgOneMinusXTe crystals the formation of helical surface states
        becomes possible due to the fact that the proper choice of  the amount of Cadmium in the alloy yields an inversion of the band ordering, see e.g.~\cite{Orlita2014,Teppe2016,Rigaux1980}. Furthermore, in materials with low $x$ the band structure can be changed from normal to inverted band ordering simply by a variation of the temperature, $T$~\cite{Teppe2016}. Such  $x$- and $T$-driven band inversions give rise to a large variety of novel physical concepts including the  Veselago lenses~\cite{Betancur-Ocampo2017a}, development of long wavelength lasers with suppression of the Auger processes~\cite{Morozov2017,Utochkin2019}, etc.

        In all publications discussing the band structure, transport phenomena, opto-electronic effects, and magneto-optical properties of  \CdXHgOneMinusXTe films the point group symmetry of the crystal is considered to be T$_d$. This follows from the crystallographic structure of the system. Here, investigating terahertz radiation induced photogalvanic currents in \CdXHgOneMinusXTe films with an inverted band structure we surprisingly observed a well pronounced circular photogalvanic effect (CPGE)~\cite{Ivchenko2005,Ganichev2005}, whose prerequisite is gyrotropy. Consequently, by symmetry arguments it is forbidden in the non-gyrotropic T$_d$ group.
        Therefore, the CPGE observation indicates either (i) the symmetry reduction of the bulk material or (ii) excitation of the photocurrent in the topological surface states formed in the material with low Cadmium concentration. We performed a careful study of terahertz radiation induced CPGE in bulk \CdXHgOneMinusXTe films with structure compositions similar to that in most of the previous studies of topological surface states in this material. We show that an attempt to ascribe the generation of the CPGE  solely to the helical topological surface states fails, because pronounced CPGE is also detected in crystals with $x$ above the critical one, which, correspondently, are characterized by a normal band ordering. To explain the origin of the observed CPGE, we suggest that the studied epitaxial films are strained and the actual symmetry of the crystal is reduced. In strained zinc-blende-type crystals, CPGE may emerge~\cite{Lyandageller1989}. We develop a microscopic theory of the CPGE for the Drude-like indirect optical transitions in bulk crystals induced by terahertz radiation, which describes well the experimental findings. We show that the radiation helicity sensitive photocurrent stems from the interference of virtual transitions via the conduction and valence bands contributing to the real optical transitions.

    \section{Samples and methods}
      \label{samples_methods}
      \begin{figure}
      	\centering
      	\includegraphics[width=\linewidth]{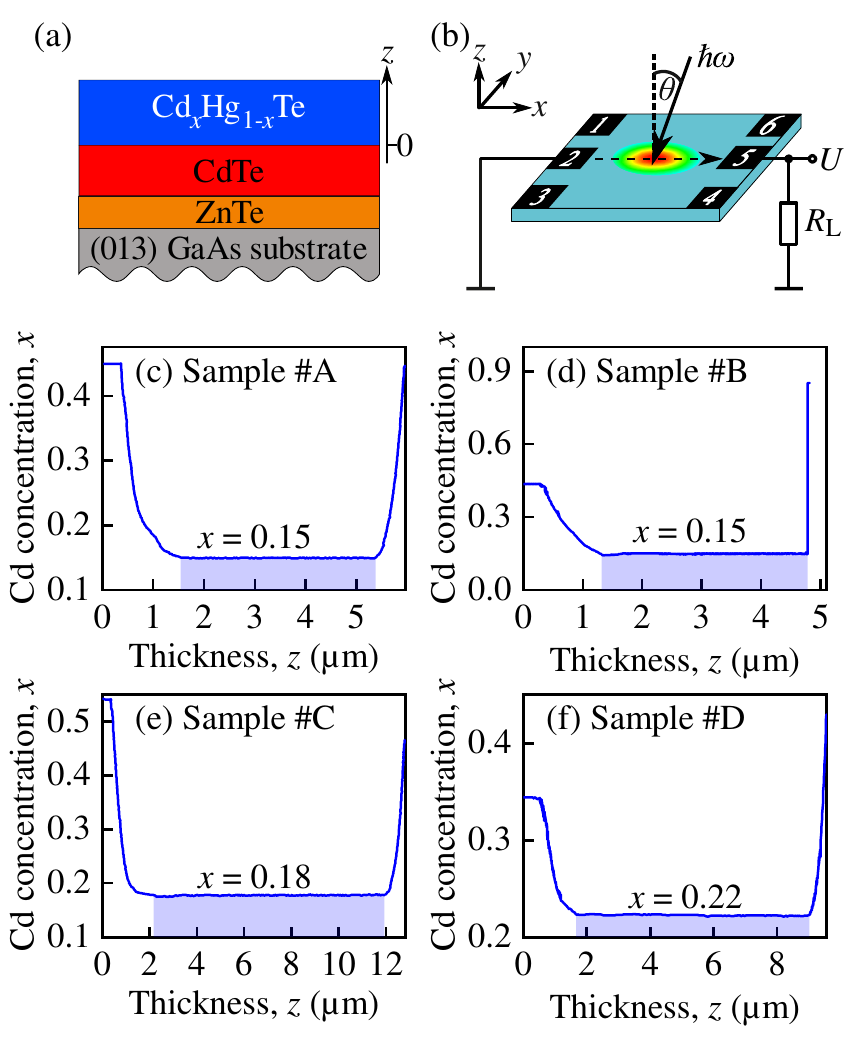}
      	\caption{\sloppy(a) Sketch of the structure composition (not to scale) (b) Experimental setup. (c)-(f) Content of Cadmium in the epitaxial layer as a function of the distance from the CdTe layer's top. The values of the Cd content $x$ in the active \CdXHgOneMinusXTe layers for samples \#A-\#D are given in the plots.}
      	\label{fig1}
      \end{figure}

      We studied \CdXHgOneMinusXTe layers with graded band gap layers at absorber boundaries grown by molecular beam epitaxy on semi-insulating (013)-oriented \ce{GaAs} substrates. \ZnTe (\NanoMeter{30} thick) and \CdTe (\MicroMeter{6} thick) buffer layers \cite{Sidorov2001} were fabricated on top of the \ce{GaAs} substrates. The structure composition is shown in Fig.~\ref{fig1}\,(a). The samples were quadratically shaped
      with approximate dimensions of $\num{5}\times\MilliMeterSquared{5}$, see Fig.~\ref{fig1}\,(b). Four samples with different profiles of the Cadmium concentration $x$ were prepared, see Fig.~\ref{fig1}\,(c)-(f). In three of these samples (\#A,\#C, and \#D) the active layers with constant $x$ contents were sandwiched (surrounded) by regions with gradually growing Cadmium concentration.  In sample \#B, by contrast, the active layer was capped by \NanoMeter{30} \ce{Cd_{0.85}Hg_{0.15}Te}. Note that due to the Fermi level position free carriers are located in the active layer and, in the films with $x<x_c$ in the topological surface states.
      The samples were supplied with several indium-soldered contacts bonded to a chip holder, see Fig.~\ref{fig1}\,(b).
      
      As a radiation source for the photocurrent measurements two types of molecular gas THz lasers were used: On the one hand, a continuous wave ($cw$) laser with a power $P$ of up to \MilliWatt{60}~\cite{Olbrich2011}, on the other hand a pulsed laser with a pulse duration of about \NanoSecond{100}, a repetition rate of \Hertz{1} and a peak power $P$ of up to \KiloWatt{60}~\cite{Ganichev2002a,Weber2008,Drexler2012}. The lasers emitted frequency lines in the range between 0.6 and \TeraHertz{2.6}. The nearly gaussian-shaped beam, controlled by pyroelectric camera~\cite{Ganichev1999},
      was focused onto the sample using a parabolic mirror. The beam spot diameter varied, depending on the radiation frequency, from 1.5 to \MilliMeter{3}. The initially linearly polarized radiation was modified applying  lambda-quarter plates made from crystal quartz. The rotation of these plates by the angle $\varphi$ with respect to the initial laser polarization plane resulted in the controllable variation of the radiation polarization state. By that the degree of circular polarization $P_{\text{circ}}$ was changed according to
      \begin{equation} \label{circ}
          P_{\rm circ} = \frac{I^{\sigma^+}-I^{\sigma^-}}{I^{\sigma^+}+I^{\sigma^-}}=\sin \left(2\varphi\right)\,,
      \end{equation}
      where $I^{\sigma^+} (I^{\sigma^-})$ is the intensity of the right- (left-) handed circularly polarized radiation part.
      The Stokes parameters defining the degrees of linear polarization were varied according to~\cite{Belkov2005}
      \begin{equation} \label{lin}
          P_{\text{\,L1}} = \frac{\cos(4\varphi)+1}{2}\,,\: P_{\text{\,L2}} = \frac{\sin(4\varphi)}{2}\,.
      \end{equation}
      THz radiation was applied at normal incidence or at oblique incidence with an angle of incidence $\theta$. The photocurrent was measured as a voltage drop across a load resistor, see Fig.~\ref{fig1}\,(b), or as the voltage drop over the sample itself.  The measurements were carried out in an optical cryostat which allowed us to access a temperature range from liquid helium to room temperature.

    \section{Results}
        \label{results}

        \begin{figure}
          	\centering
          	\includegraphics[width=\linewidth]{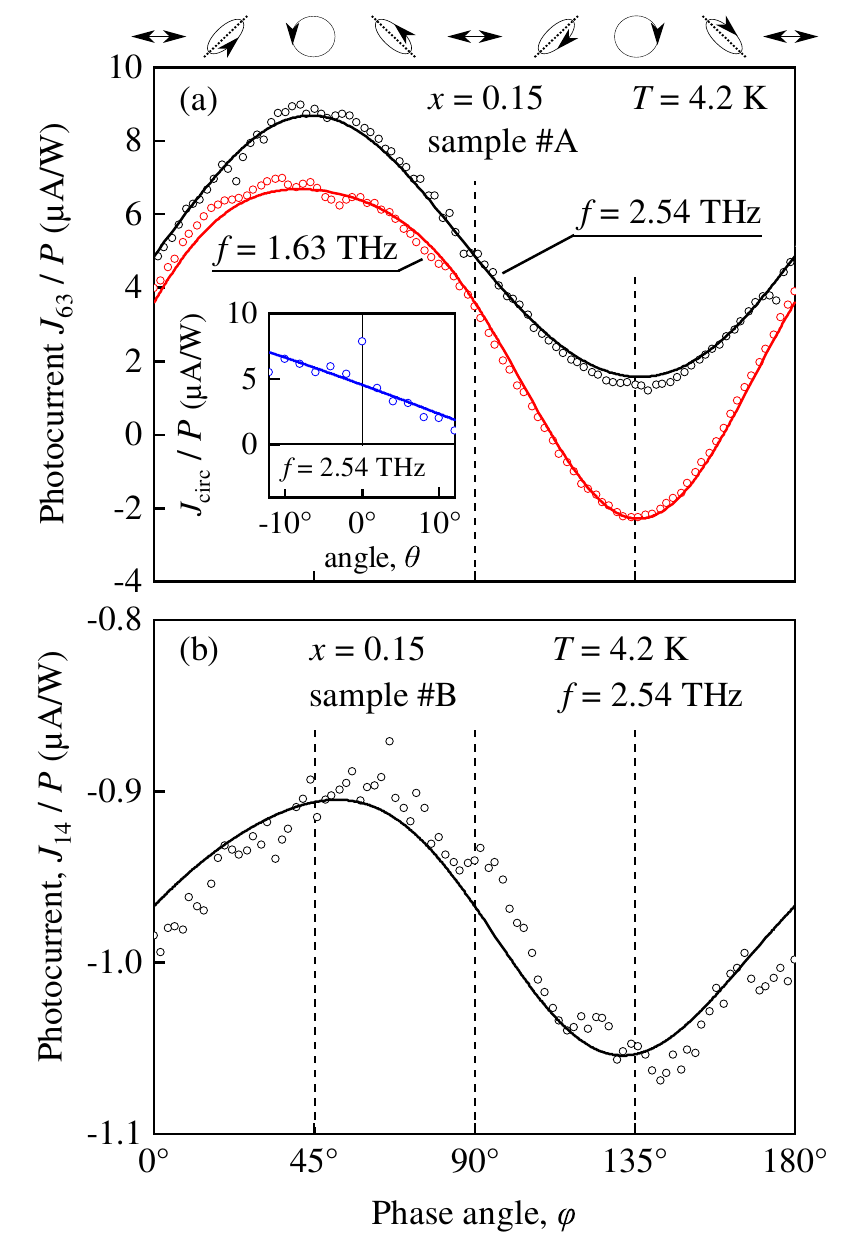}
         	\caption{\sloppy
                      Photocurrents between contacts 6 and 3 (a) and 1 and 4 (b) [see Fig.~\ref{fig1}\,(b)] as a function of radiation helicity measured at liquid helium temperature in \CdXHgOneMinusXTe crystals with a Cadmium concentration $x\approx 0.15$.  The photocurrent was excited by THz radiation from a low power $cw$ laser ($P\approx \MilliWatt{47}$) operating at frequencies 1.63 and \TeraHertz{2.54}. Solid curves show fits after Eq.~(\ref{N1}). Fitting parameters are: (a): Black curve: $J_c/P = \MicroAmperePerWatt{3.6}$, ${J_0/P= \MicroAmperePerWatt{5.1}}$,
                      ${J_{\text L 1}/P =\MicroAmperePerWatt{-0.3}}$, $J_{\text L 2}/P =\MicroAmperePerWatt{0.15}$.
                      Red curve: $J_c/P=\MicroAmperePerWatt{4.5}$,
                      $J_0/P=\MicroAmperePerWatt{2.2}$, $J_{\text L 1}/P =\MicroAmperePerWatt{1.4}$, $J_{\text L 2}/P =\MicroAmperePerWatt{0.26}$. (b): $J_c/P = \NanoAmperePerWatt{74}$, $J_0/P=\NanoAmperePerWatt{980}$, $J_{\text L 1}/P =\NanoAmperePerWatt{13}$, $J_{\text L 2}/P =\NanoAmperePerWatt{-12}$. The inset in panel (a) shows the dependence of the circular photocurrent $J_{\rm circ}$ on the angle of incidence $\theta$. The solid curve is a fit according to Eq.~(\ref{theta}). Along the top the polarization ellipses corresponding to key phase angles $\varphi$ are sketched.
          	         }
          	\label{fig2}
        \end{figure}

        The circular photocurrent was first observed at liquid helium temperature in samples \#A and \#B with $x=0.15$ being characterized by the inverted band structure. Figure~\ref{fig2} shows the data obtained at normally incident radiation of low power $cw$ radiation with frequencies 1.63 and \TeraHertz{2.54}. The overall polarization dependences can be well fitted by
        \begin{eqnarray}
            \label{N1}
            J = J_c \sin (2\varphi) + J_0+J_{\rm L1} \frac{\cos(4\varphi)+1}{2} +J_{\rm L2} \frac{\sin(4\varphi)}{2}\,.\phantom{~~~~}
        \end{eqnarray}
        In both samples the total photocurrent is dominated by the circular photocurrent described by the first term in Eq.~\eqref{N1} proportional to the coefficient $J_c$, and the polarization independent offset, $J_0$. Contributions proportional to the degrees of linear polarization, while being present, are substantially smaller. Generally, in non-centrosymmetric bulk \CdXHgOneMinusXTe crystals polarization independent photocurrents as well as those proportional to the degree of linear polarization can stem from linear photogalvanic or photon drag effects~\cite{Ivchenko2005,Ganichev2005,Sturman1992}. These effects are well known for other non-centrosymmetric materials and are out of scope of our paper.
        By contrast, in such crystals the circular photocurrent is forbidden by symmetry arguments. Therefore, below we focus on the origin and properties of this  photocurrent. Analyzing the circular photocurrent further we applied radiation at oblique incidence, see inset in Fig.~\ref{fig2}\,(a). This measurements demonstrated that the circular current $J_{\text{circ}} = [J(\varphi = \Degree{45}) - J(\varphi = \Degree{135})]/2$ varies with the angle of incidence $\theta$ as
        \begin{eqnarray}
            \label{theta}
            J_{\text{circ}} &=& J_c \cos(\theta)  +  J_c^\prime \sin(\theta)\:,
        \end{eqnarray}

        \begin{figure}
          	\centering
          	\includegraphics[width=\linewidth]{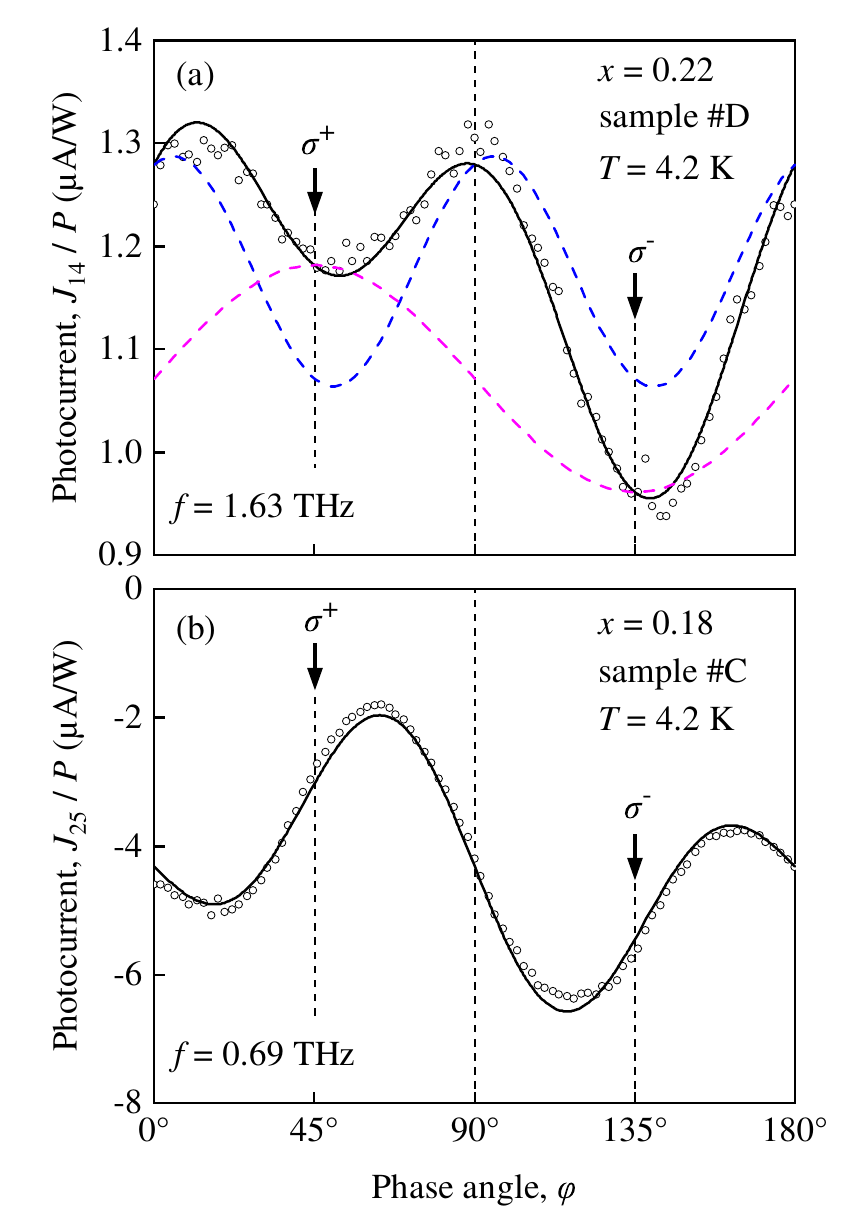}
          	\caption{\sloppy
                      Photocurrents between contacts 1 and 4 (a) and 2 and 5 (b) [see Fig.~\ref{fig1}\,(b)] as a function of radiation helicity measured at liquid helium temperature in \CdXHgOneMinusXTe crystals with active layer Cadmium concentrations $x= 0.22$ (sample \#D, $f=\TeraHertz{1.63}$) and 0.18 (sample \#C, $f=\TeraHertz{0.69}$). The photocurrent was excited by THz radiation from a low power $cw$ laser ($P\approx\MilliWatt{20}$ for $f=\TeraHertz{0.69}$; $P\approx\MilliWatt{56}$ for $f=\TeraHertz{1.63}$). The solid curves show fits according to Eq.~(\ref{N1}). Fitting parameters are: (a): $J_c/P = \MicroAmperePerWatt{0.1}$, $J_0/P=\MicroAmperePerWatt{1.1}$, $J_{\text L 1}/P =\NanoAmperePerWatt{208}$, $J_{\text L 2}/P =\NanoAmperePerWatt{82}$. (b): $J_c/P = \MicroAmperePerWatt{1.2}$, $J_0/P=\MicroAmperePerWatt{-4.2}$, $J_{\text L 1}/P =\NanoAmperePerWatt{-77}$, $J_{\text L 2}/P =\MicroAmperePerWatt{-2.8}$. The magenta dashed curve in panel (a) shows the corresponding contribution of the circular photocurrent $J_c$, the blue dashed line the one proportional to the degree of linear polarization $P_{\text L2}$. Note that both dashed lines are shifted by an offset of $J_0$.
          	        }
          	\label{fig3}
        \end{figure}

\begin{figure}
          	\centering
          	\includegraphics[width=\linewidth]{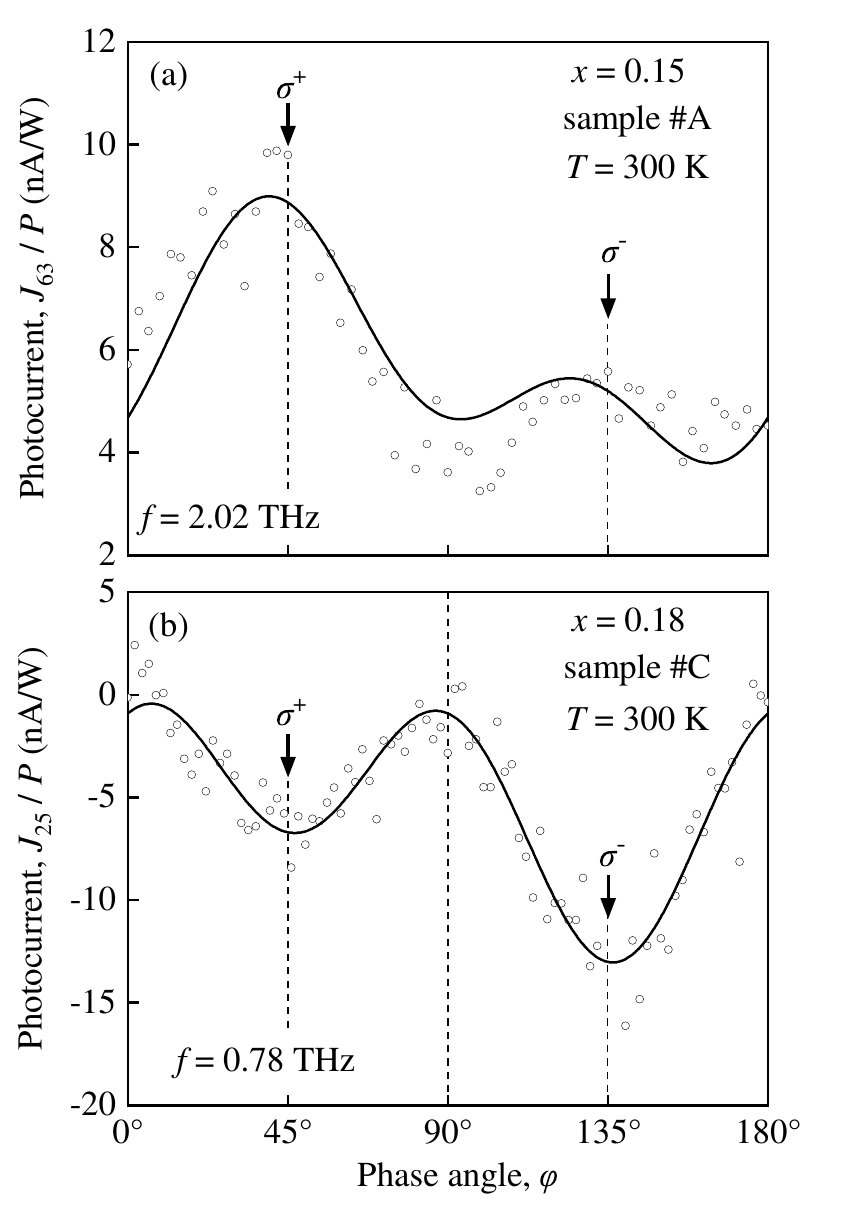}
          	\caption{\sloppy
                      Photocurrents between contacts 6 and 3 (a) and 2 and 5 (b) [see Fig.~\ref{fig1}\,(b)] as a function of radiation helicity measured at room temperature in \CdXHgOneMinusXTe crystals with an active layer Cadmium concentration $x= 0.15$ (sample \#A, $f=\TeraHertz{2.02}$) and 0.18 (sample \#C, $f=\TeraHertz{0.78}$). The photocurrent was excited by THz radiation from a high power pulsed laser ($P\approx\KiloWatt{4}$). The solid curves show fits according to Eq.~(\ref{N1}). Fitting parameters are: (a): $J_c/P = \NanoAmperePerWatt{1.8}$, $J_0/P= \NanoAmperePerWatt{7}$, $J_{\text L 1}/P=\NanoAmperePerWatt{-2.3}$ , $J_{\text L 2}/P =\NanoAmperePerWatt{1.3}$; (b):  $J_c/P= \NanoAmperePerWatt{3.2}$, $J_0/P=\NanoAmperePerWatt{-9.9}$, $J_{\text L 1}/P=\NanoAmperePerWatt{9}$, ${J_{\text L 2}/P =\NanoAmperePerWatt{1}}$.                 
                    }
          	\label{fig4}
        \end{figure}

        Samples \#A and \#B at liquid helium temperature have a \ce{Cd} concentration in the active layer lower than the critical one and, therefore, are characterized by an inverted band ordering. The latter results in the formation of topological surface states. These states, at least for samples with an abrupt increase of the Cadmium concentration $x$, like in the top layer of sample \#B, are two-dimensional and, therefore, are characterized by reduced symmetry. Consequently, in such states the circular photogalvanic effect becomes possible.
         To prove that formation of the surface states is an unambiguous requirement for the generation of the circular photocurrent in the bulk \CdXHgOneMinusXTe crystals we carried out measurements at liquid helium temperature in samples with normal band ordering (sample \#D, active layer $x=0.22$) and  with almost linear dispersion (sample \#C, active layer with Cadmium concentration $x=0.18$ being close to the critical one). Figure~\ref{fig3} shows the data obtained applying normally incident radiation of a $cw$ laser operating at frequencies 0.69, 1.63, and \TeraHertz{2.54}. In both cases the photocurrent can be well fitted by Eq.~(\ref{N1}), surprisingly, with substantial contribution of the circular photocurrent. These results already rule out band inversion and topological states formation as a prerequisite  of the CPGE in bulk \CdXHgOneMinusXTe crystals. Moreover, applying the radiation of high power pulsed lasers, which increases the sensitivity of the method, we observed that the circular photocurrent can clearly be detected even at room temperature. This is shown in Fig.~\ref{fig4} for samples \#A and \#C with $x\approx 0.15$ and $x\approx 0.18$. Previous studies of \CdXHgOneMinusXTe crystals demonstrated that at room temperature all our samples are characterized by a normal band order and no topological states are present.

        Summarizing the experimental part, our experiments provide clear evidence for the generation of the circular photogalvanic effect in bulk \CdXHgOneMinusXTe crystals with both inverted and normal band orderings, as well as for samples with critical Cadmium concentration characterized by an almost linear energy dispersion. The fact that the CPGE is clearly detected for samples with $x$ larger than the critical one demonstrates that the CPGE generation is not limited to  the films with topological surface states.

    \section{Theory and discussion}
        \label{discussion}

        \subsection{Symmetry analysis}

            Bulk \CdXHgOneMinusXTe has nominally zinc-blende crystal structure which is described by the $T_\text{d}$ point group. Despite the fact that the group lacks the center of space inversion, it is non-gyrotropic and does not support CPGE~\cite{Ivchenko2005}. However, if the spatial symmetry of the crystal is reduced further, CPGE may emerge. The most likely origin of the symmetry reduction in our samples is strain stemming, e.g., from lattice mismatch at the \CdXHgOneMinusXTe film interfaces.            
            \footnote{ Note that in samples with $x<x_c$ the CPGE caused by the optical transitions in the surface states may contribute to the total current. A similar current has been previously observed in HgTe 2D TIs \cite{Dantscher2017} and  mechanisms are overviewed in Ref.~\cite{Durnev2019} }.

            In strained zinc-blende crystals, CPGE can occur~\cite{Lyandageller1989}. To first order in strain, the dependence of the CPGE current density $\bm j$ on the static strain tensor $\bm u$ is described by
		{
	 \medmuskip=0mu
		\begin{align}
			j_{x'}= [\chi_1 (u_{y'y'}-u_{z'z'})  \hat{e}_{x'} + \chi_2 (u_{x'y'}  \hat{e}_{y'}-u_{x'z'} \hat{e}_{z'}) ] I P_{\rm circ} ,\nonumber\\
			j_{y'}= [\chi_1 (u_{z'z'}-u_{x'x'})  \hat{e}_{y'} +  \chi_2 (u_{y'z'} \hat{e}_{z'}-u_{x'y'} \hat{e}_{x'})] I P_{\rm circ}  ,\nonumber\\
			j_{z'}= [\chi_1 (u_{x'x'}-u_{y'y'})  \hat{e}_{z'} + \chi_2 (u_{x'z'}  \hat{e}_{x'}-u_{y'z'}  \hat{e}_{y'})] I P_{\rm circ} , \label{j_phen}
		\end{align} 
	}
	where $\hat{\bm e} = \bm{q}/q$ is the unit vector pointing along the photon wave vector $\bm{q}$, $I$ is the local intensity of radiation,
            %$\bm E$ is the (complex) amplitude of the ac electric field of the radiation, $i [\bm E \times \bm E^*] = |\bm E|^2 P_{circ} \, \hat{\bm o}$,
            and $x' \parallel[100]$, $y' \parallel[010]$, and $z' \parallel[001]$ are the cubic axes. The parameters $\chi_1$ and $\chi_2$ are linearly independent and describe the contributions to the photocurrent caused by normal and shear strain, respectively~\cite{Lyandageller1989}. The CPGE current vanishes in the case of hydrostatic strain that does not disturb the crystal symmetry.  Phenomenological Eqs.~\eqref{j_phen} can be readily constructed using the theory of group representations.
            In (013)-grown structures experimentally studied in our work, the tensor of strain induced by the lattice mismatch has four non-zero components $u_{xx}$, $u_{yy}$, $u_{zz}$, and $u_{yz}$ in the coordinate frame $x \parallel[100]$, $y \parallel[03\overline{1}]$, and $z \parallel[013]$ relevant to the structure orientation. Typically, the components $u_{xx}$ and $u_{yy}$ at the interfaces are determined by the lattice mismatch between the film and the buffer layer and are equal to each other. The other components, $u_{zz}$ and $u_{yz}$, can be found by minimizing the elastic energy, see, e.g., Ref.~\cite{Dantscher2015}.

%{\color{magenta}
	In the experiment on (013)-grown samples described above, we study the photocurrent excited by normally incident radiation ($\theta=0$). For this geometry, $\hat{\bm e} \parallel z$ and Eqs.~\eqref{j_phen} take the form
	 {
		\medmuskip=-0.5mu
	\begin{align}
		\label{phen_cur}
		j_y& = \left[ (\chi_1+\chi_2)(u_{zz}-u_{yy})\dfrac{\sin 2\phi}{2}+\chi_2 u_{yz} \cos 2\phi \right]  I P_{\rm circ} \:,\nonumber\\
		j_z& = -\chi_1 u_{yz} \sin 2\phi \:I P_{\rm circ}\:,
	\end{align}
}
	where $\phi$ is the angle between $[001]$ and $[013]$, $\phi=\arctan(1/3)$. A substantial photocurrent excited at normal incidence and sensitive to the degree  of circular polarization $P_{\rm circ}$ has been detected in all samples for all frequencies and temperatures used, see Figs.~\ref{fig2}-\ref{fig4} and $J_c$ in Eq.~\eqref{theta}. This photocurrent corresponds to the in-plane component of photocurrent  $j_y$  in Eq.~\eqref{phen_cur}. Note that the geometry of the samples allows us to measure only the in-plane component of the current.
\\

In addition to the helicity driven photocurrent at normal incidence experiment shows that tilting the light by an angle of incidence $\theta$ results in an additional photocurrent proportional to the degree  of circular polarization $P_{\rm circ}$ being proportional to $\sin {\theta}$, see inset in Fig.~\ref{fig2}(a). To describe the photocurrent at oblique incidence of radiation one should also take into account all components of $\hat{\bm{e}}$. For $\hat{\bm{e}}=(\sin\theta \cos \vartheta,\sin\theta \sin\vartheta,\cos \theta)$ in the $x,y,z$ coordinate frame Eqs.~\eqref{j_phen} transforms into
	\begin{align}
		j_x &= -\chi_1 \sin\theta  \cos \vartheta [(u_{zz}-u_{yy})\cos 2\phi+2 u_{yz}\sin 2\phi] I P_{\rm circ} \:,\nonumber\\
		j_y& = \biggl\{\cos\theta\left[ (\chi_1+\chi_2)(u_{zz}-u_{yy})\dfrac{\sin 2\phi}{2}+\chi_2 u_{yz} \cos 2\phi \right]  
		\nonumber\\&+\chi_1 \sin \theta \sin \vartheta
		[(u_{zz}-u_{yy})\cos 2\phi-u_{yz}\sin 2 \phi]
		\biggr\}I P_{\rm circ} \:,\nonumber\\
		j_z& = 
		-\biggl\{
		\cos \theta \;\chi_1 u_{yz} \sin 2\phi+
		\sin \theta \sin\vartheta \biggl[\chi_2 u_{yz} \cos 2 \phi  +\nonumber\\&
		-(\chi_1-\chi_2)(u_{zz}-u_{yy})  \dfrac{\sin 2\phi}{2}\biggr]
		\biggr\}
		I P_{\rm circ}\:,
	\end{align}
	here $\vartheta$ describes position of the plane of incidence  with respect to the crystallographic axes, which in the experiment is unknown. One can clearly see that even for arbitrary $\vartheta$  contributions of the photocurrent proportional to $\sin \theta$ are present, which corresponds to $J_c^\prime$ in Eq.~\eqref{theta} describing experimental data in the inset in Fig.~\ref{fig2}(a).

        \subsection{Microscopic theory}

            \begin{figure}[h]
            	\centering
            	\includegraphics[width=\linewidth]{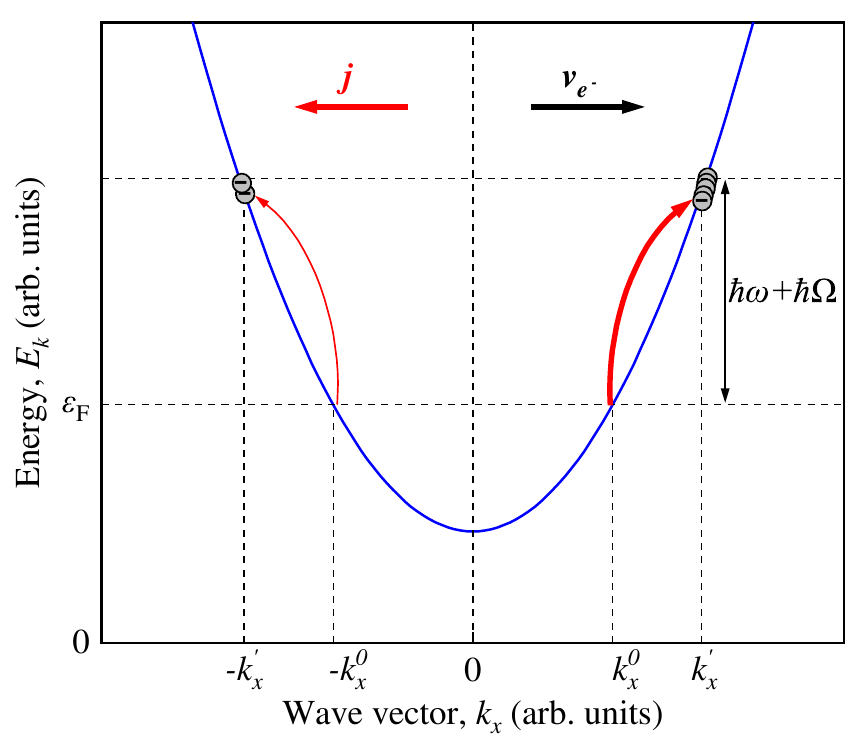}
            	\caption{\label{currentgen}\sloppy
                    		Photocurrent generation via Drude-like indirect optical transitions. The radiation with frequency $\omega$ is absorbed with simultaneous absorption or emission of a phonon with frequency $\Omega$. For circularly polarized radiation, the transitions are asymmetric in the $\bm k$-space (thin and thick arrows) resulting in a photocurrent.
                    		The asymmetry comes from interference of different pathways contributing to the transitions, see Fig.~\ref{intraband}.
              }

            \end{figure}

            Now we turn to the microscopic mechanism of the photocurrent generation. We consider \CdXHgOneMinusXTe with the content of Hg below the critical value of the transition to a 3D topological insulator. The samples have the
            conventional band structure with the $\Gamma_6$ conduction band and the $\Gamma_8$ valence band. The band gap is larger than the photon energy of THz radiation and free carriers are present in the sample. Therefore, the radiation is absorbed via indirect optical transitions (Drude-like) in the conduction band, see Fig.~\ref{currentgen}. These transitions are assisted by the scattering of electrons by phonons or static defects of the structure to simultaneously satisfy the laws of energy and quasi-momentum conservation. Indirect optical transitions are described by the second-order perturbation theory involving virtual processes via intermediate states.
            The matrix element of the real transition from the initial state $i = (\bm k, s)$, where $\bm k$ is the wave vector and $s$ is the spin index, to the final state $f = (\bm k', s')$ is given by the sum of the compound matrix elements of the virtual transitions via all possible intermediate states $j$,
            \begin{equation}\label{M_compound}
                M_{fi} = \sum_{j} \left( \frac{V_{f j} R_{j i}}{E_i - E_j} + \frac{R_{f j} V_{j i}}{E_i - E_j} \right) \,,
            \end{equation}
            where $V_{fj}$ and $R_{ji}$ are the matrix elements of the electron scattering and electron-phonon interaction,
            respectively, and $E_j$ is the total energy of the system in the $j$ state.

            \begin{figure}
            	\centering
            	\includegraphics[width=\linewidth]{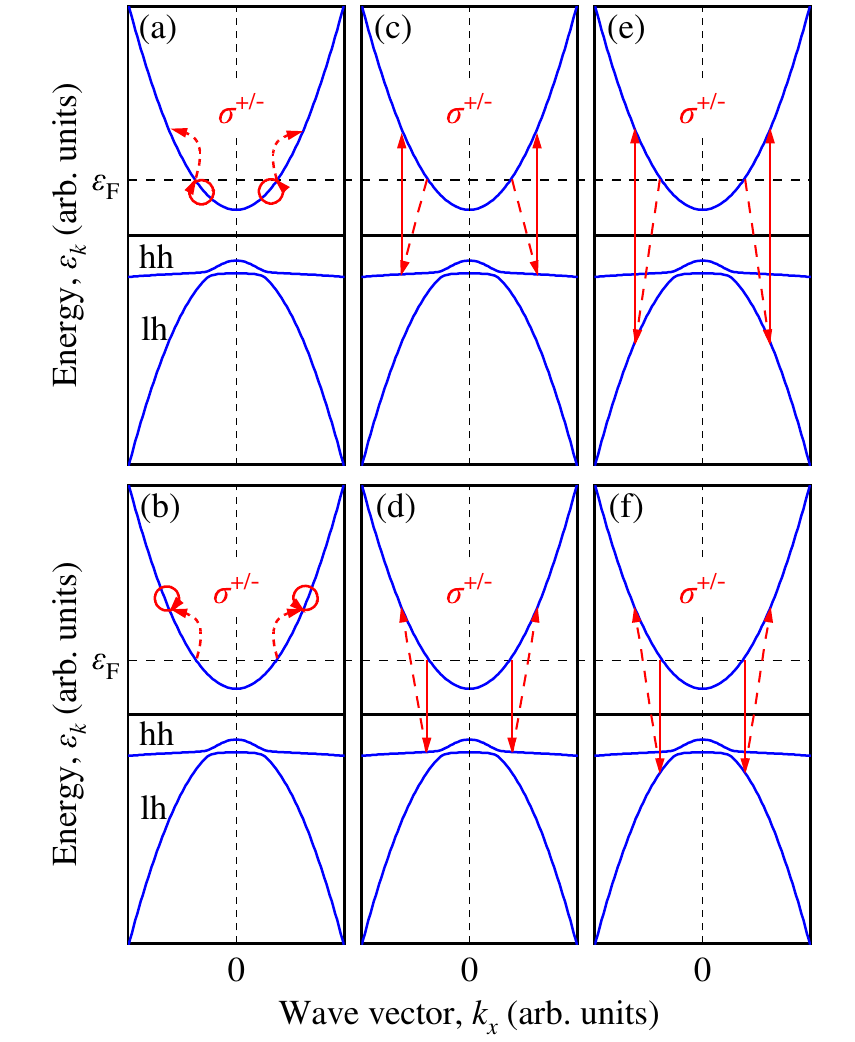}
            	\caption{\sloppy
                  			\label{intraband}
                    		Panels (a) and (b): Virtual intraband optical transitions with intermediate states in the conduction band.		 Red circles and red dashed arrows denote electron-photon interaction and electron scattering, respectively.
                        Panel (a) corresponds to the process when photon absorption is followed by electron scattering, while
                        panel (b) sketches the process in the opposite order.
                    		Panels (c)-(f): Virtual intraband optical transitions with intermediate states in the heavy-hole and light-hole subbands. Red solid and red dashed arrows denote interband electron-photon interaction and electron scattering, respectively. The mixing of heavy-hole and light-hole states by static strain, essential for the emergence of the CPGE in zinc-blende crystals, is shown as the distortion of the valence-band spectrum.
            		}
            \end{figure}

            The main contribution to radiation absorption comes from the virtual transitions with intermediate states in the conduction band, Fig.~\ref{intraband}. There are two types of such processes: the processes where electron-photon interaction is followed by electron scattering [shown in Fig.~\ref{intraband}\,(a)] and the processes with the opposite order [shown in Fig.~\ref{intraband}\,(b)]. The virtual transitions via the conduction-band states describe well the Drude absorption.
            However, they are not sensitive to the circular polarization of the radiation and do not introduce asymmetry in the electron distribution in the $\bm k$ space.

            To obtain the photocurrent, one should also take into account the virtual transitions with intermediate states in the valence band~\cite{Tarasenko2007}. Figures~\ref{intraband}\,(c)-\ref{intraband}\,(f) sketch four possible processes of such type via the heavy-hole and the light-hole bands.
            Due to the selection rules for interband optical transitions, these processes are sensitive to the radiation helicity~\cite{Ivchenko2004}.
            Moreover, their contributions can be quite large because the band gap in \CdHgTe samples, and correspondingly the denominator in Eq.~\eqref{M_compound},  is small.

            The probability of the real transitions $(\bm k, s) \rightarrow (\bm k' ,s')$ is determined by the squared modulus of the matrix element
            \begin{equation}\label{w_gen}
                \left| M_{\bm k' s', \bm k s} \right|^2 = \left| M_{\bm k' s', \bm k s}^{(c)} + M_{\bm k' s', \bm k s}^{(v)} \right|^2  \,,
            \end{equation}
            where $M_{\bm k' s', \bm k s}^{(n)}$ are the matrix elements of the virtual transitions via the $n$ band.
            It contains the interference term $2 {\rm Re} [M_{\bm k' s', \bm k s}^{(c)*} M_{\bm k' s', \bm k s}^{(v)}] $. The term does not vanish in non-centrosymmetric crystals and is responsible for the circular photogalvanic effect~\cite{Tarasenko2007}.

            The circular photocurrent emerges only in strained crystals. Therefore, in the calculation of the matrix elements of the virtual transitions we also take into account the mixing of the states by the static strain. Otherwise, the photocurrent vanishes in agreement with the symmetry {analysis} presented above. The strain-induced mixing of the heavy-hole and light-hole states is schematically shown in Fig.~\ref{intraband} as a distortion of the valence-band spectrum.

            To summarize,
            the microscopic model takes into account two ingredients essential for the CPGE: (i) the lack of space inversion center in the crystal which enables the interference of the optical transition pathways via the conduction and valence bands and (ii) the strain-induced mixing of the states.

            We calculate the photocurrent in the 6-band Kane model considering the static strain of the crystal and the electron scattering by acoustic phonons. In the basis of the $\Gamma_6$ and $\Gamma_8$ states, the Kane
            Hamiltonian has the form~\cite{Bir1974}
            \begin{equation}
                \label{KaneH}
                H =
                \begin{pmatrix}
                    0_{2} &  H_{\text{cv}}\\[0.1cm]
                    H_{\text{cv}}^\dagger & -E_g I_4
                \end{pmatrix}\,,
            \end{equation}
            where $0_{2}$ is the $2 \times 2$ zero matrix, $I_{4}$ is the $4\times 4$ identity matrix,
            \begin{equation}
                H_{\text{cv}}^\dag = P
                \begin{pmatrix}
                    -\dfrac{k_{x'}-i k_{y'}}{\sqrt{2}}  & 0 \\[0.4cm]
                    \sqrt{\dfrac{2}{3}} k_{z'} & -\dfrac{k_{x'}-i k_{y'}}{\sqrt{6}} \\[0.4cm]
                    \dfrac{k_{x'}+ i k_{y'}}{\sqrt{6}} & \sqrt{\dfrac{2}{3}} k_{z'}  \\[0.3cm]
                    0 & \dfrac{k_{x'}+ i k_{y'}}{\sqrt{2}}
                \end{pmatrix} \,,
            \end{equation}
            and $P$ is the Kane parameter.
            The Hamiltonian~\eqref{KaneH} describes six eigenstates: the conduction-band states
            $|e, \bm k, \pm 1/2 \rangle$ with the dispersion $\varepsilon_{c,k} = \hbar^2 k^2 /(2m^*)$, the states in the light-hole subband $|lh, \bm k, \pm 1/2 \rangle$ with the dispersion $\varepsilon_{lh,k} = -E_g - \hbar^2 k^2 /(2m^*)$, and the dispersionless states in the  heavy-hole subband $|hh, \bm k, \pm 3/2 \rangle$ with the energy $\varepsilon_{hh} = -E_g$. Here, $m^* = 3\hbar^2 E_g/(4P^2)$ is the effective mass.
            The Hamiltonian of electron-photon interaction is given by
            \begin{equation}
                R = - \frac{e}{\hbar c} \bm A \cdot \nabla_{\bm k} H \,,
            \end{equation}
            where $e$ is the electron charge, $c$ is the speed of light, $\bm A$ is the vector potential of the electromagnetic field related to the radiation intensity $I$ by $I = A^2 \omega^2 n_{\omega}/(2 \pi c)$, and $n_{\omega}$ is the refractive
            index of the crystal.
            The strain Hamiltonian in the 6-band model is given by
            \begin{equation}
                \label{strainH}
                V =
                \begin{pmatrix}
                    \Xi_c\,  \operatorname{Tr}\left(\bm \epsilon\right)\, I_2\, &  V_{\text{cv}}\\[0.1cm]
                    V_{\text{cv}}^\dagger & V_{\text{BP}}
                \end{pmatrix} \,,
            \end{equation}
            where $\Xi_c$ in the conduction-band deformation potential, $\bm \epsilon$ is the strain tensor,
            $V_{\text{BP}}$ is the Bir-Pikus Hamiltonian which, in the spherical approximation, has the form~\cite{Bir1974}
            \begin{equation}
                V_{\text{BP}} =
                 \left(a+\dfrac{5}{4}b\right)
                I_4 \operatorname{Tr}\left(\bm \epsilon\right) -
                b \sum_{\alpha\beta} J_\alpha J_\beta \,\epsilon_{\alpha\beta} \,,
            \end{equation}
            $a$ and $b$ are the valence-band deformation potentials, $J_{\alpha}$ are the matrices of the angular momentum $3/2$, and $V_{\text{cv}}$ is the part describing the strain-induced coupling of the
            $\Gamma_6$ and $\Gamma_8$ states in zinc-blende crystals~\cite{Pikus1988,Ivchenko2004},
            \begin{equation}
                V_{\text{cv}}^\dag=\Xi_{\text{cv}}
                \begin{pmatrix}
                    -\dfrac{i \epsilon_{y'z'} + \epsilon_{x'z'}}{\sqrt{2}}  &  0 \\[0.4cm]
                    i\sqrt{\dfrac{2}{3}} \epsilon_{x'y'} & -\dfrac{i \epsilon_{y'z'} + \epsilon_{x'z'}}{\sqrt{6}} \\[0.4cm]
                    \dfrac{i \epsilon_{y'z'}- \epsilon_{x'z'}}{\sqrt{6}} & i\sqrt{\dfrac{2}{3}} \epsilon_{x'y'} \\
                    0 &  \dfrac{i \epsilon_{y'z'}- \epsilon_{x'z'}}{\sqrt{2}}
                \end{pmatrix} \,,
            \end{equation}
            and $\Xi_{cv}$ is the interband deformation potential. The Hamiltonian~\eqref{strainH} is used to calculate both the mixing of the states by the static strain $\bm u$ and the electron scattering by longitudinal acoustic (LA) phonons.
            The tensor of strain produced by the LA phonons is given by
            \begin{equation}
                \epsilon_{\alpha\beta} = i \sum_{\bm q} \frac{q_{\alpha} q_{\beta}}{q^2} \sqrt{\frac{\hbar \, q^2}{2 \rho\, \Omega_q}} \left( \text{e}^{i \bm q \cdot \bm r} a_{\bm q} - \text{e}^{- i \bm q \cdot \bm r} a_{\bm q}^\dag \right) \,,
            \end{equation}
            where $\bm q$ is the phonon wave vector, $\rho$ is the crystal density, $\Omega_q = c_s q$ is the photon frequency, $c_s$ is the speed of longitudinal sound,  and $a_{\bm q}$ and $a_{\bm q}^\dag$ are the operators of phonon annihilation and creation, respectively. We assume that the photon frequency $\omega$ considerably exceeds both the frequency of phonons involved in scattering $\Omega_{|\bm k' - \bm k|}$ and the scattering rate $\tau^{-1}$, and that the phonon occupation numbers are large, i.e., $k_B T/ (\hbar \Omega_{|\bm k' - \bm k|}) \gg 1$, where $T$ is the temperature. To the first order in the wave vector, the matrix elements of the virtual transitions via the conduction band with the absorption of a photon and the simultaneous emission ($+$) or absorption ($-$) of a LA phonon has the form
            \begin{equation}
                \label{m0}
                M^{(c,\pm)}_{\bm k' s', \bm k s} = \pm i\,   \dfrac{e \, m^* \, \Xi_c \, \bm{A}(\bm{k}-\bm{k}') }{c \, \omega} \sqrt{\dfrac{k_B T}{2\, \rho\, c_s^2}} \,  \, \delta_{s',s} \,.
            \end{equation}
            %
            %These transitions are spin conserved.

            In calculating the matrix elements of the virtual transitions via the valence band we take into account the mixing of the heavy-hole and light-hole stated by the static strain. The corresponding contribution to $M^{(v)}_{\bm k' s', \bm k s}$
            proportional to the static strain can be obtained by the third-order perturbation theory and is given by
            \begin{widetext}
                \begin{multline}
                    \label{m1}
                    M^{(v,\pm)}_{\bm k' s', \bm k s} = \sum_{m,n} \biggl\{
                    \dfrac{V_{\bm k' s', \bm k m}^{(\pm)}  U_{\bm k m,\bm k n} R_{\bm k n, \bm k s}}
                    {(\varepsilon_{c k}+\hbar \omega-\varepsilon_{mk})(\varepsilon_{ck} +\hbar \omega - \varepsilon_{nk})}
                    + \dfrac{R_{\bm k' s', \bm k' m} U_{\bm k' m, \bm k' n}  V^{(\pm)}_{\bm k' n, \bm k s}}
                    {(\varepsilon_{ck} - \varepsilon_{m k'})(\varepsilon_{ck} - \varepsilon_{n k'})} \\
                    + \dfrac{V^{(\pm)}_{\bm k' s', \bm k m}  R_{\bm k m,\bm k n}  U_{\bm k n, \bm k s}}
                    {(\varepsilon_{ck} + \hbar \omega - \varepsilon_{m k})(\varepsilon_{ck} - \varepsilon_{n k})}
                    + \dfrac{U_{\bm k' s, \bm k' m} V^{(\pm)}_{\bm k' m, \bm k n} R_{\bm k n, \bm k s}}
                    {(\varepsilon_{ck} + \hbar\omega - \varepsilon_{m k}')(\varepsilon_{ck}+\hbar \omega - \varepsilon_{n k})} \\
                    + \dfrac{U_{\bm k' s,\bm k' m} R_{\bm k' m,\bm k' n} V^{(\pm)}_{\bm k' n,\bm k cs}}
                    {(\varepsilon_{ck}+\hbar \omega - \varepsilon_{nk'})(\varepsilon_{ck} -\varepsilon_{mk'})}
                    + \dfrac{R_{\bm k' s, \bm k' m} V^{(\pm)}_{\bm k' m, \bm k n} U_{\bm k n, \bm k s}}
                    {(\varepsilon_{ck} - \varepsilon_{m k'})(\varepsilon_{ck} - \varepsilon_{nk})}
                    \biggr\} \,,
                \end{multline}
            \end{widetext}
            where the indexes $m$ and $n$ run over the valence subbands. The matrix elements of electron scattering with the emission or absorption of a LA phonon $V_{\bm k' s', \bm k m}^{(\pm)}$ and the matrix elements of heavy-hole--light-hole mixing by the static strain $U_{\bm k m,\bm k n}$ are calculated using the strain Hamiltonian~\eqref{strainH}. The exact analytical expression for $M^{(v)}_{\bm k' s', \bm k s}$ is too cumbersome to be printed. An estimation capturing the dependence on the band-structure parameters and deformation potentials is the following
            $M^{(v)}_{\bm k' s, \bm k s} \sim  (e \, \omega\, u\, b\, A\, P\, \Xi_{cv} / c \,E_g^3) \sqrt{k_B T /\rho\, c_s^2}$.

            The photocurrent in the relaxation time approximation is given by the standard expression
            \begin{multline}
                \label{current}
                \bm{j} =  e\tau \sum_{\bm{k} \bm{k}' s, \pm} \dfrac{2\pi}{\hbar} |M_{\bm k' s, \bm k s}^{(c,\pm)}
                + M_{\bm k' s, \bm k s}^{(v,\pm)}|^2 \\
                \times  (\bm{v}_{\bm k'} - \bm{v}_{\bm k})
                [f(\varepsilon_{ck}) - f(\varepsilon_{ck'})] \delta(\varepsilon_{ck'} - \varepsilon_{ck} - \hbar \omega)\,,
            \end{multline}
            where $\tau$ is the momentum relaxation time, $\bm v = \hbar \bm k/m^*$ is the electron velocity, and $f(\varepsilon_{ck})$ is the equilibrium Fermi-Dirac distribution function.

          	Labour-consuming calculation of Eq.~\eqref{current} with the matrix elements~\eqref{m0} and~\eqref{m1} yields
          	the CPGE current~\eqref{j_phen} with the coefficients
          	\begin{multline}
            		\label{chi_eq}
            		\chi_1 = -\frac{512 \pi^2}{315} \dfrac{ e^3 \tau}{\hbar^2 c\,  n_\omega}  \dfrac{k_B T}{\rho\, c_s^2}
            		\dfrac{\Xi_{\text{cv}}\, \Xi_{\text{c}}\, b }{(\hbar\omega)^2} \dfrac{P^5}{E_g^5} \\
            		\times \sum_{\bm{k}\bm{k}'}
            		[f(\varepsilon_{ck}) - f(\varepsilon_{ck'})]   (k'^2+k^2)\,
            		\delta(\varepsilon_{ck'} - \varepsilon_{ck} - \hbar \omega)
          	\end{multline}
          	and $\chi_2 = -(5/3) \chi_1$.
			Now using Eq.~\eqref{phen_cur} one can already obtain the equation for the in-plane photocurrent which describes well all major experimental findings.
          	To proceed further, we assume that the momentum relaxation of electrons is also determined by the electron-phonon interaction. For quasi-elastic scattering by LA phonons, the momentum relaxation time depends on energy and
          	has the form
          	\begin{equation}
              	\frac{1}{\tau(\varepsilon_k)} = \frac{m^*}{\pi \hbar^3} \frac{k_B T \, \Xi_c^2}{\rho \,c_s^2}  \sqrt{\frac{2m^* \varepsilon_k}{\hbar^2}} \,.
          	\end{equation}
          	The relaxation time of the average electron velocity in the model of the drifting electron gas is given by
          	\begin{equation}
              	\frac{1}{\tau} = \sum_{\bm k} \frac{\varepsilon_k}{\tau(\varepsilon_k)} \frac{\operatorname{d} f(\varepsilon_k)}{\operatorname{d} \varepsilon_k} /
              	\sum_{\bm k} \varepsilon_k \frac{\operatorname{d} f(\varepsilon_k)}{\operatorname{d} \varepsilon_k} \,,
          	\end{equation}
          	which gives
          	\begin{equation}\label{tau_LA}
          	     \frac{1}{\tau} = \dfrac{4 m^*}{3\pi \hbar^3} \frac{k_B T \, \Xi_c^2}{\rho \,c_s^2} \, \bar{k} \,,
          	\end{equation}
          	where \begin{equation}
          	\bar{k} = \sum_{\bm k} |\bm k| f(\varepsilon_k) / \sum_{\bm k} f(\varepsilon_k)
          	\end{equation} 
          	is the mean value of $|\bm k|$.
          	Taking into account Eq.~\eqref{tau_LA} and ${m^* = 3\hbar^2 E_g/(4P^2)}$, we can rewrite Eq.~\eqref{chi_eq} in the form
          	\begin{multline}
            		\label{chi_eq2}
            		\chi_1 = -\frac{18}{35 \pi}  \dfrac{e^3}{ \hbar^3 \omega^2 \,c\, n_\omega \,\bar{k}}
            		\dfrac{\Xi_{\text{cv}}}{\Xi_{\text{c}}}  \dfrac{b}{E_g^2 P} \\
            		\times \int d\varepsilon \sqrt{\varepsilon (\varepsilon+\hbar \omega)}(2\varepsilon+\hbar \omega)
            		[f(\varepsilon) - f(\varepsilon+\hbar \omega)] \,.
          	\end{multline}

          	In the case when the photon energy $\hbar\omega$ is much less than the mean electron energy $\bar{\varepsilon}$,
          	Eq.~\eqref{chi_eq2} takes the form
          	\begin{equation}
              	\label{chi_eq3}
              	\chi_1 = -\frac{64 \pi}{35 }  \dfrac{e^3  N_e b}{ \hbar^2 \omega \, c  \,n_\omega}
              	\dfrac{\Xi_{\text{cv}}}{\Xi_{\text{c}}}\frac{P^3}{E_g^4} \,,
          	\end{equation}
          	where $N_e = 2 \sum_{\bm k} f_{\varepsilon_{\bm k}}$ is the electron density.
          	
	Finally, the in-plane photocurrent induced at normal incidence of radiation in (013)-oriented films  is given by
				\begin{multline}
				\label{final_pge_eq}
				j_y = \frac{64 \pi}{21 }  \dfrac{e^3  N_e b}{ \hbar^2 \omega  c n_\omega}
				\dfrac{\Xi_{\text{cv}}}{\Xi_{\text{c}}}\frac{P^3}{E_g^4}
				\times\\\left[\dfrac{2}{5}(u_{zz}-u_{yy})\dfrac{\sin 2\phi}{2}-u_{yz} \cos 2\phi \right]  I P_{\rm circ} \:.
				\end{multline}

\begin{figure}
	\centering
	\includegraphics[width=\linewidth]{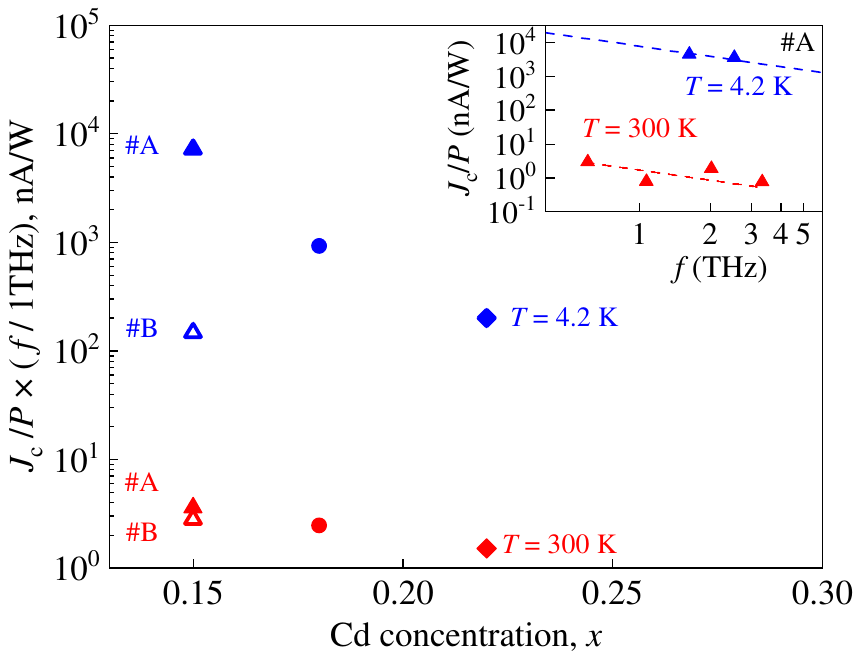}
	\caption{\sloppy Dependence of the circular photocurrent contribution normalized to radiation frequency on the cadmium concentration $x$ for temperatures $T=\SI{4.2}{\kelvin}$ (blue symbols) and  $T=\SI{300}{\kelvin}$ (red symbols). The inset shows the dependence of the circular photocurrent on the frequency for sample \#A at temperatures $T=\SI{4.2}{\kelvin}$ (blue symbols) and  $T=\SI{300}{\kelvin}$ (red symbols). Dashed lines in the inset indicate a fit of $J_c/P$ according to $J_c/P\propto 1/f$.
	}
	\label{fig7}
\end{figure}
          	This equation describes the observed polarization dependence of the photocurrent, its almost linear increase with the frequency decrease as well as drastic increase of the photocurrent magnitude by cooling the sample from room to liquid helium temperature. Indeed, the proportionality of the  photocurrent $J_c$ to the radiation helicity $P_{\rm circ}$ has already been addressed above and is clearly seen in Figs.~\ref{fig1}-\ref{fig4} obtained for 4.2 and 300~K. The inset in Fig.~\ref{fig7} presenting the circular photocurrent as a function of the radiation frequency shows that it varies after $J_c \propto 1/\omega$, which agrees with Eq.~\ref{final_pge_eq}. Results obtained for samples with different Cd contents $x$ and temperatures, are characterized by substantially different energy gaps. According to Eq.~\ref{final_pge_eq} the latter should strongly affect the circular photocurrent magnitude yielding drastic increase of the current amplitude upon the energy gap reduction ($J_c\propto 1/{E_g^4}$). Figure~\ref{fig7} and Tab.~\ref{pge_table} present the magnitude of the circular photocurrent  obtained for different samples and temperatures. Note that to make comparison of data obtained at different frequencies possible, we used the fact that $J_c~\propto 1/\omega$ and normalized the data to the radiation frequency. The figure reveals that at fixed temperature, either 300~K or 4.2~K, $J_c/P$ substantially reduces with the band gap increase, see Table~\ref{pge_table}. Furthermore, comparing the photocurrent magnitudes in each individual sample we see that the photocurrent is increased by more than two orders of magnitude at the temperature decrease from 300 to 4.2~K. This fact is in agreement with substantial reduction of the band gap at low temperatures. The values of $E_g$ for different sample compositions $x$ and temperatures are taken from Refs.~\cite{Teppe2016,Rigaux1980,Laurenti1990}. We note that for samples \#A and \#B at \SI{4.2}{\kelvin}, the band gap becomes negative and applicability of Eq.~\ref{final_pge_eq} is not justified. Now we estimate from the room temperature experiment the strain  $u = \left[\dfrac{2}{5}(u_{zz}-u_{yy})\dfrac{\sin 2\phi}{2}-u_{yz} \cos 2\phi \right]$. Room temperature data are used because, under these conditions, all the samples have a normal band dispersion, thus no topologically protected surface states can be formed and this case is relevant to the above derived equation for the photocurrent. The current density $j_c$ was obtained from the photocurrent $J_c$ measured in experiment after $j_c/I = J_c/P\cdot S_{\rm beam}/(d\cdot d_{\rm beam})  $, where $d$ is the width of the conducting channel, $d_{\rm beam}$ is the radiation beam diameter and $S_{\rm beam}$ is the beam area. Note that while for room temperature results this calculation seems to be reasonable at helium temperatures it is not straightforward, because free carriers and, consequently, the current may be  distributed inhomogeneous across the sample. The latter unknown factor makes a quantitative comparison of \SI{4.2}{\kelvin} data with the theory difficult. Table \ref{pge_table} presents the strain $u$ obtained using the parameters $2 m_0 (P/\hbar)^2 = 18.8$~eV \cite{Novik2005},  $b = -1.4$~eV, and $n_{\omega} = 4.6$ \cite{Adachi2004}. While the rate $\Xi_{cv}/\Xi_c$ is unknown for HgTe, we use $\Xi_{cv}/\Xi_{c}=0.3$ for GaAs \cite{Pikus1988}. Table~\ref{pge_table} shows that the strain $u$ in samples with different compositions varies from $1.6\times10^{-7}$  to   $6.9\times10^{-6}$ and is well below the strain $u_{max}=1.52\times10^{-3}$ estimated by minimizing the elastic energy for pure HgTe ($x=0$) deposited on \ce{CdTe} substrate. \cite{Dantscher2015}. We note that a small value of the strain is not surprising because we deal with thick films. Under this condition strain is expected to be $z$-coordinate dependent and be strongest at the bottom boundary. 	
          \begin{table*}
	\begin{tabular}{ccc|ccccc|c}
		
		Sample	&	$x$        &\makecell{$f$\\ (\TeraHertz{})}&\makecell{$J_c/P$\\(\NanoAmperePerWatt{})}& \makecell{$j_c/I$\\(\MicroAmperePerWatt{})}& \makecell{$N$\\(\SI{}{\centi\meter\tothe{-3}})}&  \makecell{$E_g$\\(\SI{}{\milli\electronvolt})}&\makecell{$u$\\$(10^{-6})$}&\makecell{$E_g$\\(\SI{}{\milli\electronvolt})}\\
		&&&&&$T=\SI{300}{\kelvin}$&&&$T=\SI{4.2}{\kelvin}$\\
		\hline
		\hline 
		\rule{0pt}{3ex}\#A&0.15&0.60&2.9&2.1&$1.8\times10^{17}$&83&0.4&-32\\
		\#B&0.15   & 2.03  &1.4& 1.0 &$1.3\times10^{17}$& $83$& 0.5&-32\\   
		\#C&0.18  & 0.78  &0.8& 0.8 &$1.1\times10^{17}$& $123$&1.6&20\\		    			
		\#D&0.22   & 2.03  &0.1& 0.2&$4.1\times10^{16}$& $182$&6.9&94  \\    			
	\end{tabular} \\
          		\caption{Experimental data for normalized photocurrent, carrier densities, band gaps and the calculated strain for samples \#A-\#D at \SI{300}{\kelvin} and additionally the band gap values for \SI{4.2}{\kelvin} \cite{Laurenti1990}.}
          		          		\label{pge_table}
  	\end{table*}
            Finally we note that other possible sources of the CPGE in the films made of non-gyrotropic crystals are (i) other (unrelated to strain) mechanics of bulk symmetry reduction; (ii) interface related effects, such as 2D electron states, which may occur at the interfaces or anisotropic scattering of bulk electrons at the interfaces; and (iii) topological surface states for samples with $x<x_c$. 
             A photocurrent sensitive to the helicity of incident photons can also emerge as a result of the circular-to-linear polarization conversion in a birefringent medium and the linear photogalvanic effect (LPGE). This scenario can be excluded since the birefringence in the THz range is very weak while the detected CPGE and LPGE currents are comparable.

    \section{Summary}
        \label{summary}
        To summarize, we demonstrate in our work the symmetry breaking in \ce{CdHgTe} structures resulting in a helicity-sensitive photogalvanic current with opposite directions for excitation with right- and left-handed circularly polarized radiation. The circular photocurrent is present in films with different Cadmium concentrations as well as in a wide temperature range, which supports the conclusion of the strain-induced symmetry reduction. The developed theoretical model describes the experimental data in second-order perturbation theory considering the interference of matrix elements in the probability of indirect Drude-like optical transitions involving virtual processes via intermediate states.

    \section{Acknowledgments}
        \label{acknow}

        The support from the CENTERA, DFG priority program SFB 1277 (project A04), the Elite Network of Bavaria (K-NW-2013-247), the Volkswagen Stiftung Program, and the Program N13 of the Presidium of the RAS is gratefully acknowledged.

        I.\,Y. thanks the National Science Centre, Poland, (grant No. UMO-2017/25/N/ST3/00408) for support. G.V.B. also acknowledges support from the "BASIS" foundation.

    \bibliography{ref}

\end{document}